\documentclass[prb,twocolumn,superscriptaddress,amsmath,amssymb, showkeys]{revtex4-2}

\usepackage{amssymb,amsmath}
\usepackage{color}
\usepackage{graphicx}
\usepackage{dcolumn}
\usepackage{bm}
\usepackage{latexsym,epsfig}
\usepackage{hyperref}

\usepackage{marvosym}
\usepackage[caption=false]{subfig}

\usepackage{amsmath, amssymb, graphics, setspace}
\usepackage{graphicx}
\usepackage{dcolumn}
\usepackage{bm}
\usepackage{lipsum}
\usepackage{tikz}

\newcommand{\cEQ}[1]{Eq.~\ref{#1}}
\newcommand{\cFIG}[1]{Fig.~\ref{#1}}

\newcommand{\commentB}[1]{\textcolor{black}{{#1}}}
\newcommand{\mic}{\,$\mu$m\:}

\begin{document}

\preprint{APS/123-QED}

\title{\commentB{Bilayer Ion Trap Design for 2D Arrays}}

\author{Gavin N. Nop}
\email{gnnop@iastate.edu}
\affiliation{The Ames National Laboratory, U.S. Department
of Energy, Iowa State University, Ames, IA 50011, USA}
\affiliation{Department of Mathematics,  Iowa State University, Ames, Iowa 50011, USA}
\author{Jonathan D. H. Smith}
\email{jdhsmith@iastate.edu}
\affiliation{The Ames National Laboratory, U.S. Department
of Energy, Iowa State University, Ames, IA 50011, USA}
\affiliation{Department of Mathematics,  Iowa State University, Ames, Iowa 50011, USA}
\author{Daniel Stick}
\email{dlstick@sandia.gov}
\affiliation{Sandia National Laboratories, Albuquerque, New Mexico 87185, USA}
\author{Durga Paudyal}
\email{durga@ameslab.gov}
\affiliation{The Ames National Laboratory, U.S. Department of Energy, Iowa State University, Ames, IA 50011, USA}
\affiliation{Department of Electrical and Computer Engineering,  Iowa State University, Ames, Iowa 50011, USA}


\date{\today}

\begin{abstract}
Junctions are fundamental elements that support qubit locomotion in two-dimensional ion trap arrays and enhance connectivity in emerging trapped-ion quantum computers. In surface ion traps they have typically been implemented by shaping radio frequency (RF) electrodes in a single plane to minimize the disturbance to the pseudopotential. However, this method introduces issues related to RF lead routing that can increase power dissipation and the likelihood of voltage breakdown.
Here, we propose and simulate a novel two-layer junction design incorporating two perpendicularly \commentB{rotoreflected (rotated, then reflected)} linear ion traps. The traps are vertically separated, and create a trapping potential between their respective planes.
The orthogonal orientation of the RF electrodes of each trap relative to the other provides perpendicular axes of confinement that can be used to realize transport in two dimensions.
While this design introduces manufacturing and operating challenges, as now two separate structures have to be precisely positioned relative to each other in the vertical direction and optical access from the top is obscured, it obviates the need to route RF leads below the top surface of the trap and eliminates the pseudopotential bumps that occur in typical junctions.
In this paper the stability of idealized ion transfer in the new configuration is demonstrated, both by solving the Mathieu equation analytically to identify the stable regions and by numerically modeling ion dynamics. Our novel junction layout \commentB{has the potential to enhance} the flexibility of microfabricated ion trap control to enable large-scale trapped-ion quantum computing.
\end{abstract}

\maketitle

\section{Introduction}
High-fidelity quantum operations and engineering advances over the last decade have established trapped ions as strong candidates for constructing a practical quantum computer. The fundamental component of a trapped-ion quantum computer is the RF Paul trap, which uses oscillating and static voltages applied to electrodes to constrain ions whose internal states provide the physical basis for the logical qubits. Lasers and/or microwaves are used to initialize, read out, and perform quantum gates on the ionic qubits \cite{haffner:2008}. 
While microfabricated linear traps have been developed for over fifteen years \cite{stick:2005,seidelin:2006,leibrandt:2009,allcock:2012}, and have been used for sophisticated multi-ion experiments \cite{ionq, pino:2021}, microfabricated junction traps \cite{moehring:2011} have only recently shown sub-quantum excitation during transport \cite{quantinuum}.  Even with that demonstration, key challenges remain to the scaling up of trapped-ion arrays to achieve the
connectivity required for enlarged quantum volume, faster computational cycles, and increased qubit capacity \cite{qubit_scaling}.

An RF Paul trap confines ions at distances of tens \cite{ivory:2021} to hundreds \cite{pogorelov:2021} of microns from the closest surfaces, effectively isolating them from the environment. While Earnshaw's theorem prohibits the creation of an electrostatic potential well, an RF \commentB{voltage} can be applied to particular electrodes to form a time-averaged pseudo-potential with a minimum determined by the RF electrode geometry \cite{RF_trap_design}. Quasi-static voltages applied to separate control electrodes can be used to store and move ions along the \emph{RF null},
a line along which the RF electric field is zero and there is a resulting minimum in the radially confining pseudopotential.
The original Paul traps typically had hyperbolic electrodes \cite{paul_trap}, but modern microfabrication techniques use layered planes of materials to create two-dimensional trap geometries \cite{ion_trap_history}.
Linear RF nulls are common, but can be modified to produce curves and junctions for the transfer of ions between multiple ion traps \cite{design_criteria}. 

There are two main categories of trapped-ion quantum computing architectures: the quantum charge-coupled device \cite{kielpinski:2002}, and those that rely on stationary chains of ions with all-to-all intra-chain gate operations and remote entanglement via photons to connect separate chains \cite{brown:2016}. In the former case, ion transport is the primary conduit for entangling distant ions.  In the latter case, while photonic interconnects are used to connect distant chains, some level of ion transport between distinct but nearby chains may still be advantageous.
A 2D ion layout reduces the scaling of transport times over arbitrary distances from $O({n})$ for a 1D layout to $O({n}^{1/2})$, where $n$ is the number of ions \cite{qubit_movement}.
A 2D layout also better matches the connectivity requirements of surface codes used for quantum error correction \cite{yu:2013, quanConn}.
Both $3$-way \cite{shu:2014} and $4$-way \cite{burton:2023} junctions enable grid-based ion transport 
to support arbitrary 2D movement. However, scaling the array size using these junctions presents the challenge of islanded RF electrodes \cite{amini:2010, blain:2021} that require electrical vias and leads routed underneath the top metal surface, raising the capacitance and resistance of the device, and thereby increasing the RF power dissipation.  These buried leads also increase the likelihood of voltage breakdown between RF and ground, as they introduce more locations where the RF electrode or lead approaches a grounded electrode or lead, often within only a few microns.

Our design achieves 2D connectivity with simple rectangular RF electrodes, avoids islanded RF electrodes, and does not require RF vias. It utilizes low-excitation transport protocols already developed for use with
current linear micro-fabricated ion trap designs. The primary challenge is that two separate trap layers have to be assembled, imposing design constraints on the conventional microfabrication techniques and limiting optical access.  While these are important considerations, we consider here whether the trap, if fabricated and assembled, would form a viable junction.

The paper begins with the initial layouts and conventions for the junction design (Sec.~\ref{sec:basicDesign}), and then presents the formal treatment of ion transfer stability (Sec.~\ref{sec:analyticTreatment}, \ref{sec:computationalResults}). It concludes with a discussion of the unique challenges and opportunities that arise \commentB{at an architectural} level for the ion trap configuration (Sec.~\ref{sec:trapConfiguration}, \ref{sec:conclusions}).

\section{Basic Junction Design}\label{sec:basicDesign}

\begin{figure}[h]
\includegraphics[width=0.48\textwidth]{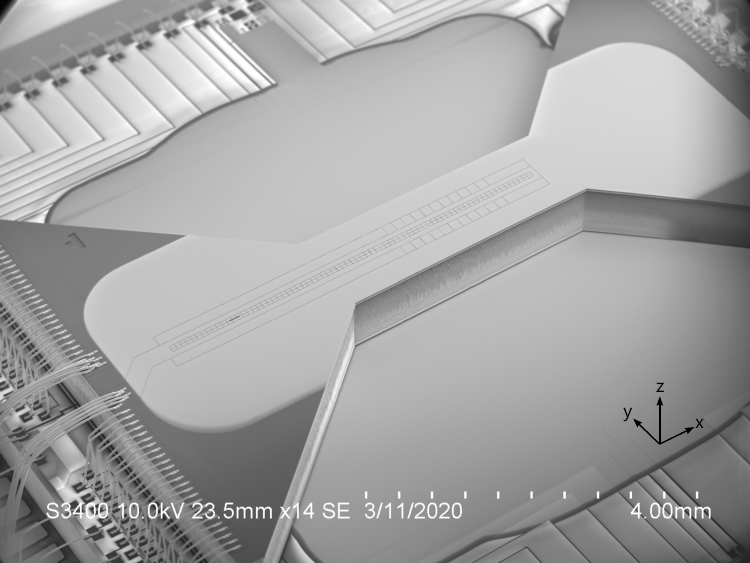}
\caption{\label{fig:singleTrap} SEM micrograph of a Peregrine trap.  This microfabricated trap confines ions to a linear region 72\mic above the surface.  The trap consists of alternating aluminum electrode and oxide insulation layers, with an evaporated gold region on top.
}
\end{figure}

Ions in a surface trap are confined at points above the top plane of the trap at the RF null.
Right-handed orthogonal coordinates are used throughout this document, whereby the ion travels along the $x$-axis, and the $z$-axis is perpendicular to the trap surface \cite{rf_null_visual}. The Peregrine trap serves as an example to illustrate the proposed junction \cite{revelle:2020}. \cFIG{fig:singleTrap} displays the entire device, where the relevant trapping portion is in the central isthmus.

\begin{figure}[h]
\includegraphics[width=0.48\textwidth]{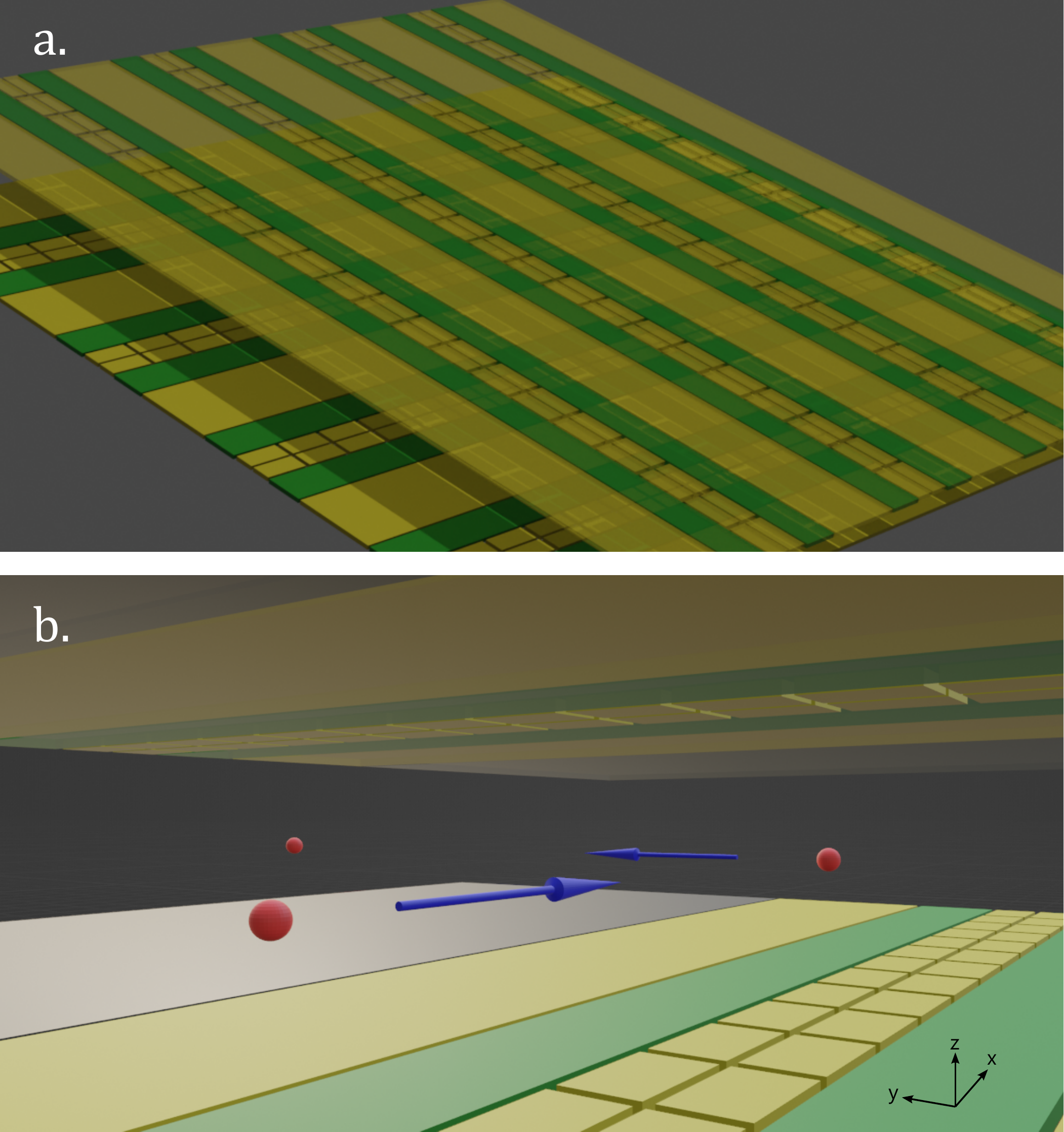}
\caption{\label{fig:dualTrap} A conceptual rendering of the two-layer junction design, here integrated into a potential configuration of parallel arrays. Two traps in a pair are vertically separated and rotoreflected relative to each other. Part a) shows a zoomed-out view that highlights the linear nature of the traps, and their orientation relative to each other.  The RF electrodes are green, and the control electrodes are yellow.  Part b) shows the position of the ions in the plane between two traps, with a junction that lies at the point of symmetry. The ions and arrows are for illustration; as explained later in the paper the concept of operation supports one of the traps providing confinement at any given time.  
}
\end{figure}

In our proposed junction design, the ion trap is duplicated, translated vertically, inverted, and then rotated $90^\circ$ about the $z$ axis, as illustrated in \cFIG{fig:dualTrap}. In any such combined configuration, the fields generated by each individual trap are modified by the second trap acting as a ground plane.
\commentB{If the RF rails on the top and and bottom traps were aligned and voltages were applied to both at the same time, then the RF null would be half-way between the trap planes (like a 4 rod trap but with additional grounded regions).}
However, with only one set of RF electrodes fully on at a time, the RF null ends up at a distance of less than half the trap separation (closer to the trap that is on), and so the two RF nulls do not directly overlap. Nevertheless, the slight separation of the RF nulls is not an obstacle to ion transfer.
For two Peregrine traps offset by 50\mic\!, the RF nulls are 23.7\mic from each surface. The simulations described later show that ions can still be transferred from one trap to the other, across the 2.6\mic vertical separation of the RF nulls.


If RF voltages were applied simultaneously to both traps, there would be a pseudopotential barrier preventing an ion from being shuttled from one trap to the other. Therefore, a scheme is employed that starts with the RF voltage applied to \commentB{trap $b$ (bottom} trap in \cFIG{fig:dualTrap}b) but not trap \commentB{$t$ (top} trap in \cFIG{fig:dualTrap}b). Once an ion is transported to the intersection of the traps, the static and RF voltages are gradually switched until the ion is confined by trap \commentB{$t$}. \commentB{In the case where the RF electrodes on one trap are all connected, all ions have to transfer from one trap to the other at the same time, or be lost. An alternative would be to segment the RF electrodes and apply independent voltages to different segments, but this would require the vias that we wanted to avoid in the first place.}

\commentB{While this design involves passing ions from one trap to another vertically and in the process turning one trap on and another off, a demonstration of passing ions from one trap to another in the same plane by abutting (but not contacting) the separate RF electrodes has been successfully achieved \cite{akhtar:2023}.  In this case the RF pseudo-potential provides nominally undisturbed confinement across chips. There has also been a demonstration of vertical transport of charged macroscopic particles in a 3D trap \cite{hucul:2008} that uses suspended wires. While this 3D geometry would face similar optical access challenges and likely greater manufacturing challenges compared to the bilayer trap described in this paper, it shows that 3D traps with motion in all three dimensions is possible.}

For modelling purposes, the $x$-, $y$- and $z$-orientations are determined by trap \commentB{$b$}. The origin of the coordinate system occurs at the nexus of the junction and is the symmetric point between the two traps.

A mathematical model describes the transfer of the ion from its trapping location in the bottom trap at $x=0$, $y=0$, $z=-s$ to the \commentB{top} trap at $x=0$, $y=0$, $z=s$. Here, $s$ is half the vertical separation between the traps.  The potential in the bottom trap is \commentB{$\phi_b(x,y,z,t) = \phi_{C,b}(x,y,z,t) + \phi_{RF, b}(x,y,z,t) =  \phi_C(x,y,z+s)  + \phi_{RF}(x,y,z+s,t)$, where $\phi_b$ is the sum of the potentials due to the control electrodes ($\phi_C$) and RF electrodes ($\phi_{RF}$). The potential arising from the top trap when the same voltages are applied to the corresponding electrodes is $\phi_t(x,y,z,t) = \phi_{C, t}(x,y,z,t) + \phi_{RF, t}(x,y,z,t) = \phi_C(y,-x,s-z)  + \phi_{RF}(y,-x,s-z,t)$.}

The model specifies a protocol for transferring the ion from the bottom trap to the top trap using a time-dependent scalar function, $f(t)$, that specifies how the voltages applied to the trap electrodes are scaled in time.  \commentB{The total potential is therefore
\begin{equation}\label{E:TotalPot}
\Phi(x,y,z,t)=(1-f(t))\phi_b + f(t)\phi_t\,.
\end{equation}}
Simple forms of the transfer function $f(t)$ could be instantaneous Heaviside functions or linear transitions from $0$ to $1$, but in practice smooth transitions to limit induced motional excitations are preferable.



The subsequent section determines the stability of the trap throughout the transition from $f(t)=0$ to $f(t)=1$ over a definite time interval.

\section{Analytic ion stability 
}\label{sec:analyticTreatment}

The analytic expression describing an ion in a Paul trap is the Mathieu equation. We begin with a treatment of the relevant aspects of the Mathieu equation in one dimension \cite{basic_mathieu_solution}, and then extend and apply it to the question of trapped-ion stability in three dimensions. 
Stability diagrams locate where confined particle motions can be maintained.
Numerical flight simulations then show how the ion is stably controlled at all points during the transfer from one trap to the other.

\subsection{Mathieu stability}

The Mathieu equation can be written \commentB{(in the ``canonical form'' of \cite{basic_mathieu_solution})} as
\begin{equation}\label{E:MathieuE}
q'' + (U + 2V\cos(2t))q = 0,
\end{equation}
with $t$ a unitless parameterization for time defined by the RF drive voltage, and $q$ the ion position in one dimension. For the current application, $U$ corresponds to the static confining voltage along one dimension. The sign of $U$ determines if the static potential is trapping or anti-trapping. The parameter $V$ corresponds to the root mean square 
magnitude of the oscillating voltage. Solutions are stable if they are locally confined as $t\rightarrow \infty$; otherwise, they are unstable.
\commentB{Although here we use the unitless $t$ for mathematical simplicity, often the Mathieu equation is instead written with $\Omega t/2$, where $\Omega$ is the drive frequency in radians/s and $t$ is the real time in seconds.}

\begin{figure}[h]
\includegraphics[width=0.5\textwidth]{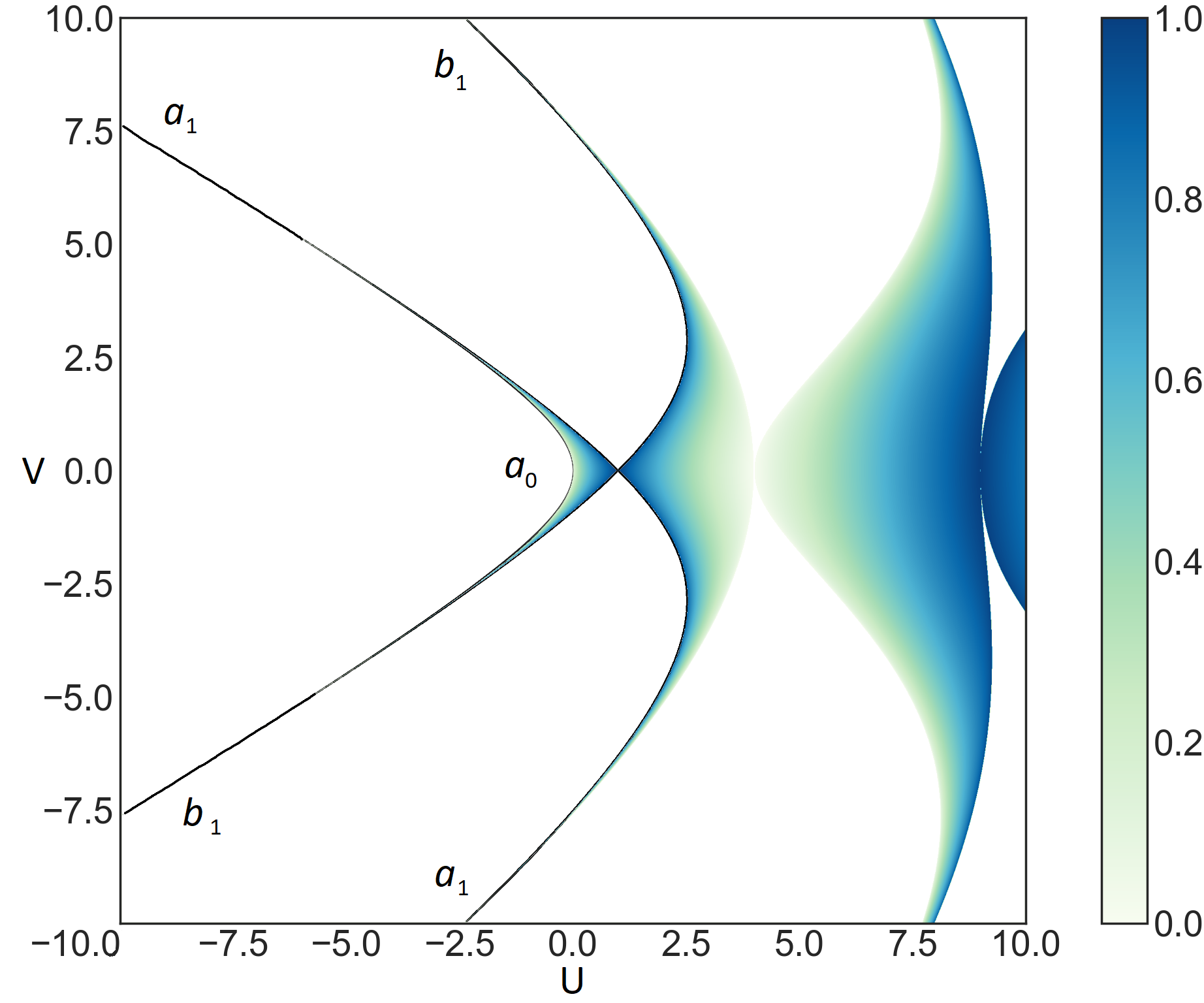}
\caption{\label{fig:singleStable} Stability diagram for parameter pairs $(U, V)$ in the one-dimensional Mathieu equation, \cEQ{E:MathieuE}. \commentB{The coloring corresponds to $w$ values in Sec.~\ref{sec:analyticTreatment} with intermediate values being more stable, and the extremes near $0.0,1.0$ being more eccentric. However all colored points correspond to stable orbits, and internal colorings simply give an idea of the qualitative character such orbits have.} The plot displays $0$ for unstable points and positive values for stable solutions. Relevant analytically smooth bounding functions $a_0(V), a_1(V), b_1(V)$ are colored black and labeled. }
\end{figure}


Taking even integers $r$, solutions of the form
\begin{equation}\label{E:MatAnstz}
q=\sum_{r\in 2\mathbb{Z}} c_{r} e^{(w + ri)t}
\end{equation}
are considered, noting that their boundedness depends on the real part of $w \in \mathbb{C}$, an arbitrary parameter. For \cEQ{E:MatAnstz} to be a solution (with non-zero constants $c_r$), the algebraic equations
$$\zeta_{r} c_{r-2} + c_{r} + \zeta_{r}c_{r+2} = 0$$
with $\zeta_{r} = 
{V}/{[(r-w i)^2 - U]}$ must be completely satisfied. Thus, the vanishing of the meromorphic function \commentB{(The ratio of two analytic functions over the complex plane.)}
$$
\Delta(iw) = 
\begin{vmatrix}
\cdots &\cdots     &\cdots    &\cdots     &\cdots    &\cdots    &\cdots\\
\cdots &\zeta_{-2}&1         &\zeta_{-2}& 0        &0         &\cdots\\
\cdots &0          &\zeta_{0}&1          &\zeta_{0}&0         &\cdots\\
\cdots &0          &0         &\zeta_{2} &1         &\zeta_{2}&\cdots\\
\cdots &\cdots     &\cdots    &\cdots     &\cdots    &\cdots    &\cdots
\end{vmatrix}
$$
over $\mathbb{C}$ locates the non-trivial solutions. 

The function $\Delta(iw)$ of $w$ is periodic, with period $2i$. Further, it is even: $\Delta(iw)=\Delta(-iw)$. Its behavior is determined by the strip $0\le\Im(w)\le 1$. The only singularities of the function $\Delta(iw)$ are simple poles occurring where $U-(r-iw)^2=0$. In particular, there is only one pole on the strip $0\le\Im(w)\le 1$. 

Next, consider the even meromorphic function
\begin{equation*}
\chi(iw)={\big[\cos \pi i w - \cos \pi \sqrt{U}\big]}^{-1} 
\end{equation*}
with period $2i$, and also with simple poles at the same locations as those of $\Delta(iw)$. It follows that there is a constant $C$, determined by the ratio of the residues of the functions $\Delta$ and $\chi$ at their common pole on the strip, such that the function
$
\Delta(iw)-C\xi(iw)
$
has no singularities, and is therefore constant by Liouville's theorem.
Setting $w=0$ determines $C$, and further algebra yields the existence of bounded solutions when 
\begin{equation}\label{eq:stabCond}
    w = \frac{1}{\pi} \cos^{-1}[1-\Delta(0)(1-\cosh \pi \sqrt{U})]
\end{equation}
for real $w$. These $w$-values may be located by iterative approximation,
readily identifying the set $\mathcal S$ of stable pairs $(U,V)$ as displayed in \cFIG{fig:singleStable} (compare Ref. 
\cite[Fig.~8(a)]{basic_mathieu_solution}). The relevant curves separating the stable and unstable regions are labelled with standard function names $a_0, a_1, b_1$\commentB{, with approximate forms available in the literature \cite{basic_mathieu_solution}}.

\subsection{\commentB{Junction stability}}
\label{sec:jstable}

The analysis of an ion trap involves three distinct Mathieu equations, one for each dimension. The interplay of these equations imposes additional stability constraints. In the process of transferring an ion from trap \commentB{$b$} to trap \commentB{$t$}, the static and RF potentials are expressed as quadratic functions at the RF null, assuming symmetric control voltages and neglecting the minute higher-order contributions: 
$$
\phi = \alpha x^2 + \beta y^2 + \gamma Z^2.
$$
Here $Z = z-s^{\prime}$ (where $s^{\prime}=-s \; (+s)$ for the lower (upper) trap) is the vertical distance from the RF null of the applicable controlling trap.
Additionally, given the long, straight RF electrodes, the RF potentials are assumed to be ideally linear along the $x$-axis for the bottom trap ($y$-axis for the top), such that $\alpha_{RF} = 0$ ($\beta_{RF} = 0$). The absence of free charges implies $\alpha + \beta + \gamma = 0$ for each field, resulting in a simplified bottom-trap control potential of:
$$
\phi_{C,b} = \alpha x^2 + \beta y^2 + \gamma Z^2
$$
with $\gamma=-\alpha-\beta$, 
and a bottom-trap RF potential of:
$$ 
\phi_{RF,b} = \cos(2 t)(-2\mu y^2 + 2\mu Z^2)
$$
with the parameter $\mu$ tracking the RF voltage. Similar equations can be generated for the top trap by interchanging the $x$ and $y$ terms.  The RF voltages applied to the bottom and top traps are set to be in-phase.

Accounting for the relative rotoreflection of the top and bottom traps, 
the total control and RF potentials as the ion is transferred from the bottom trap to the top trap (described by Eqn. \ref{E:TotalPot}) are calculated to be
\begin{equation*}
\begin{split}
    \Phi_C & = \bigl[(1-f(t))\alpha + f(t) \beta \bigr]x^2 \\
    & + \bigl[(1-f(t))\beta + f(t) \alpha \bigr]y^2 \\ 
    & + \bigl(1-f(t)\bigr) \gamma (z+s)^2 + f(t) \gamma (s-z)^2 
\end{split}
\end{equation*}
and 
\begin{equation*}
\begin{split}
    \Phi_{RF} & = \cos(2 t)\bigl[- f(t) 2\mu x^2 - \bigl(1-f(t)\bigr)2\mu y^2 \\
    & + \bigl(1-f(t)\bigr) 2\mu (z+s)^2 + f(t) 2\mu (s-z)^2 \bigr]\,.
\end{split}
\end{equation*}
Without loss of generality we set $f(t)=t/T=\tau$, where $T$ is the unitless time over which ions are transferred at a linear rate from one trap to the other. For evaluating the stability of this trap it therefore suffices to determine if the pairs
\begin{equation}\label{E:TimeStab}
\big(\tau\beta + (1-\tau)\alpha,\tau\mu\big)
\quad
\mbox{ and } 
\quad
    \big(-\alpha-\beta, \mu\big)
\end{equation}
lie in the stable set $\mathcal S$ for $0\le \tau \le 1$. By assumption, the motional excitation of the ion and the spatial separation of the traps is small enough such that, if the ion is trapped by one pseudopotential at the junction, it is also trapped by the other (although the nulls are not the same). The validity of this assumption is borne out by the flight simulations described in Sec.~\ref{sec:computationalResults}

\begin{figure}[h]
\includegraphics[width=0.5\textwidth]{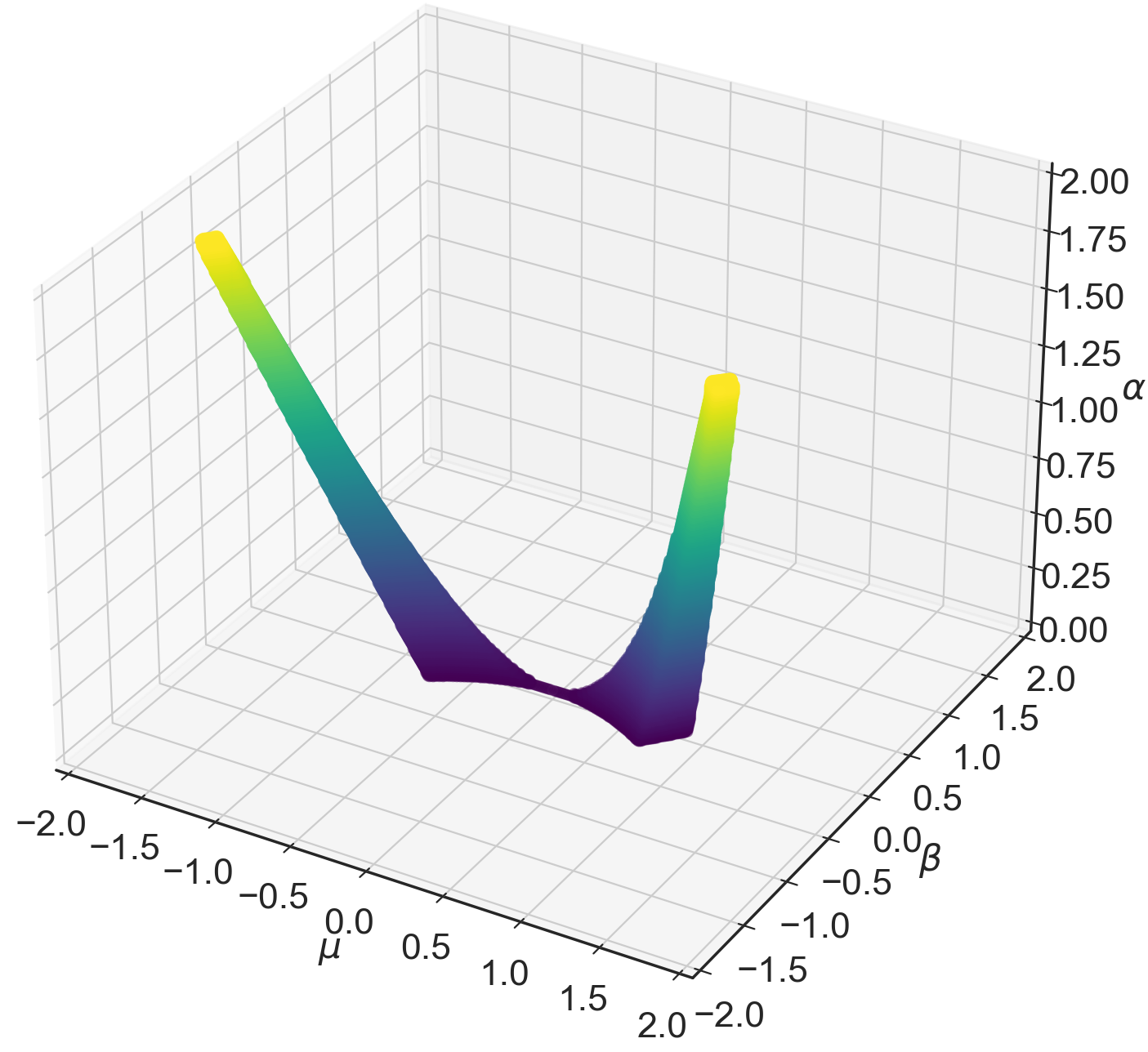}
\caption{\label{fig:simpleMap} The three-dimensional region of stability for a single linear Paul trap. The region is symmetric in the RF voltage parameter $\mu$. Since the parameter $\alpha$ controls the ion along the trap axis, it must be positive. The parameter $\beta$ tracks the static electrode transverse containment. \commentB{The height of the solid is emphasized by the coloring to make the $\alpha$-extent of the solid more visible.}}
\end{figure}

When an ion is held at a single point in a simple linear trap, stability merely requires that the three points $(\alpha, 0), (\beta, \mu)$, and $(-\alpha-\beta, \mu)$ all lie in the stable set $\mathcal{S}$. This situation, illustrated in \cFIG{fig:simpleMap}, 
applies at the start of a junction transfer, when $\tau=0$. When $\alpha = 0$ in the figure, the two stability conditions imposed are $(\beta, \mu), (-\beta, \mu)\in\mathcal S$. This corresponds to requiring the point $(\beta,\mu)$ to lie both in the region displayed in \cFIG{fig:singleStable}, and in the corresponding region obtained by reflection about the line $U=0$. The bounding surfaces of this stability region are determined by the inequalities 
\begin{align*}
a_0(\mu)<\beta\,,\ (-\alpha - \beta)<
\min\{a_1(\mu),b_1(\mu)\}
\end{align*}
using the functions defined in Fig.~\ref{fig:singleStable},
and $\alpha>0$.

\begin{figure}[h]
\includegraphics[width=0.5\textwidth]{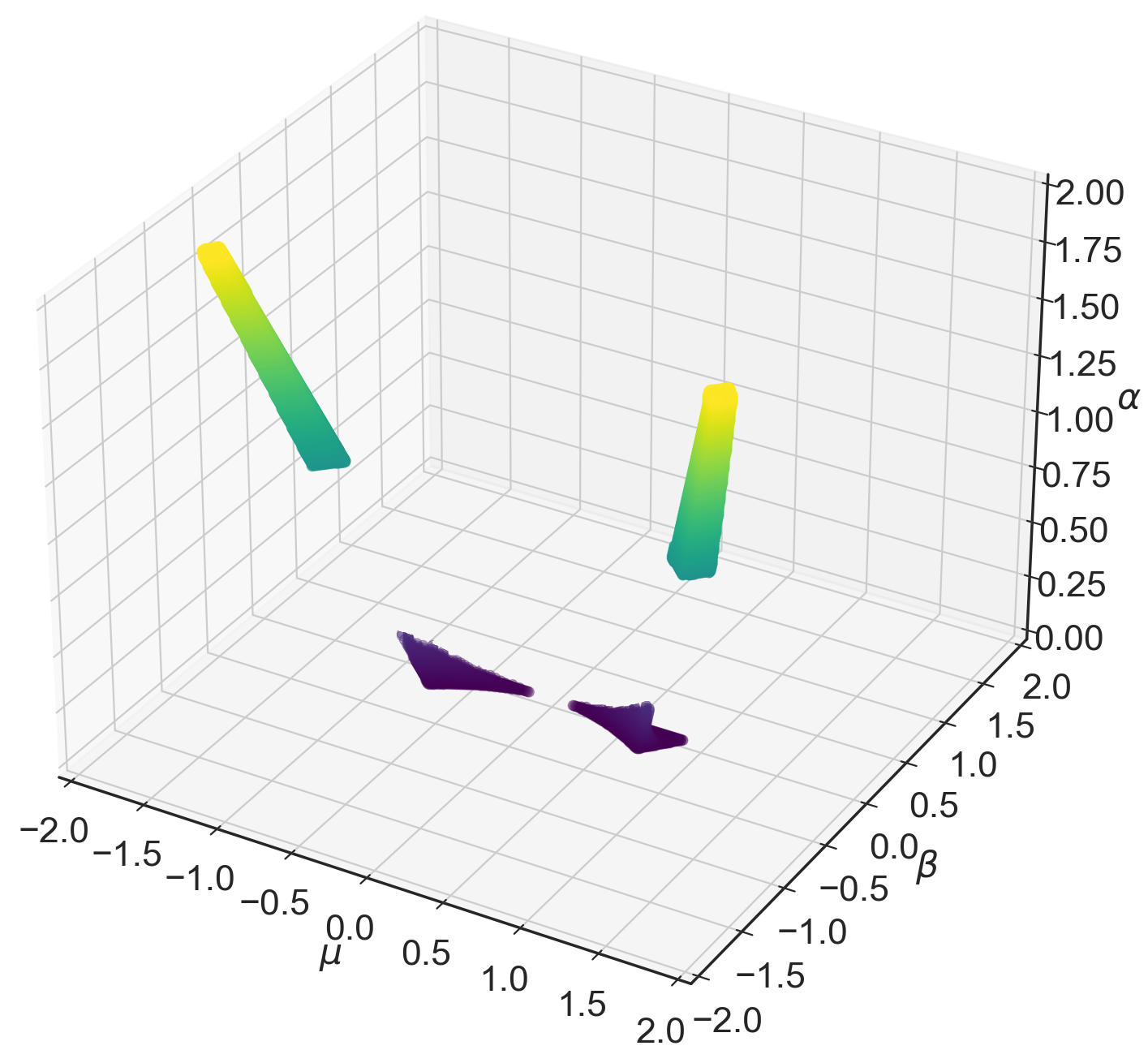}
\caption{\label{fig:complexMap} \commentB{The components of the solid for a single Paul trap in \cFIG{fig:simpleMap} that are unstable for a junction transfer. The set difference between these components is the stable region in which a junction can operate. The bounds of the regions are calculated by testing the inclusions specified in Eq.~\ref{E:TimeStab}}. A map contained within the region depicted in \cFIG{fig:simpleMap}, illustrating the parameter space where a junction between two linear Paul traps would become \emph{unstable}. The set difference between the regions depicted in \cFIG{fig:simpleMap} and here determines the stability region for the junctions. The relevant unstable regions for single ion traps are the twin dark blue volumes at the bottom. These unstable regions may be avoided by tuning the control electrodes.}
\end{figure}

A complete junction transfer $(\mu, \beta, \alpha)$ is considered to be stable
if each of the three individual one-dimensional Mathieu equations remain stable throughout the transfer operation. Thus, as described by \cEQ{E:TimeStab}, the stability condition is the conjunction of the simple stability condition that $(-\alpha-\beta, \mu)$ lies in $\mathcal{S}$, and the dynamic stability condition that $(\alpha(1-\tau) + \beta \tau, \mu \tau)$ lies in $\mathcal{S}$ for all $0\le \tau \le 1$. A map of parameter points $(\mu, \beta, \alpha)$ which are stable for a linear trap, but unstable for a junction transfer, is shown in \cFIG{fig:complexMap}. Note the four portions, all lying in the region depicted in \cFIG{fig:simpleMap}. 
The bottom portions are banned for small values of $\alpha$. These regions correspond to when the ion escapes along the $x$ axis when the line formed by $\big(\tau\beta + (1-\tau)\alpha,\tau\mu\big)$ crosses below $a_0$. The boundary of this region is then computed, considering the line $(\alpha, 0) - (\beta, \mu)$, to obtain the inequalities
$$
m = \frac{\beta - \alpha}{a_0' \mu}
\
\mbox{ and }
\
a_0(m) < \frac{\beta - \alpha}{\mu}m + \alpha\,.
$$
The top portions correspond to $\alpha > 1.0$, the point of separation between the regions of stability in \cFIG{fig:singleStable}. Physically, this region appears because $\alpha > 0$ confines the ion along the $x$-axis, but repels it along the $z$-axis, no longer confining the ion when the repelling force due to large $\alpha$ rivals the confining pseudoforce due to $\mu$. The different regions of stability are connected by a single point. When $(\alpha, 0)$ is in the second region of stability, $(\alpha(1-\tau) + \beta \tau, \mu \tau)$ intersects an unstable region, and the junction fails. A protocol holding the parameters in this region would create an unstable ion trap junction. On the other hand, protocols within the difference set between the respective regions of \cFIG{fig:simpleMap} and \cFIG{fig:complexMap} keep the junction stable. 
Qualitative analysis of ion motional frequencies  indicates that low values of $\alpha$ and $\mu$ are expected for trap operation. Thus, the $\alpha > 1.0$ region of instability does not impede ion junction operation in practice.

We have thus demonstrated analytically that transfer protocols, stable under adiabatic operation, exist for the proposed geometry represented by the set difference between \cFIG{fig:simpleMap} and \cFIG{fig:complexMap}. In those cases for which the point $(\mu,\beta,\alpha)$ lies in the region specified by \cFIG{fig:complexMap}, the control electrodes corresponding to $\alpha$ and $\beta$ can be tuned to ensure junction stability.






\section{Numerical ion stability 
}\label{sec:computationalResults}

As a complement to the analysis in the preceding section, numerical simulations were also performed to demonstrate the successful transfer of an ion from one trap to the other, under standard stability conditions for intermediate junction states. An electrostatic model was generated, and control solutions to trap the ion in all three directions at each of the final null points (corresponding to $f(t)$ = 0 and 1) were obtained.  The resulting potentials were used to calculate the electric fields and ion dynamics using the Runge-Kutta (RK4) method \cite{rk4}.  The following subsections provide more details about these steps.

\subsection{Field and flight simulation}
Each electrode generates an electric field based on the supplied voltage and trap geometry.
The geometry analyzed here uses the same lateral layout of electrodes as in the Peregrine trap, which by itself has an ion height of 72\mic\!. When two of these planes of electrodes are combined in the orientation shown in  \cFIG{fig:dualTrap} with a 50\mic separation, the RF null moves to 23.7\mic above the lower trap. For an applied RF voltage with a $31\,\mathrm{MHz}$ drive frequency and $56\,\mathrm{V}$ amplitude, this trap produces a $2.75\,\mathrm{MHz}$ radial trapping frequency for ytterbium ions, corresponding to $\mu=0.25$. The general methodology for correlating practical trap designs with the analytic treatment in this paper is explained in the caption for \cFIG{fig:traj}, which includes the unusually high value of $\mu = 0.75$ to demonstrate trap dynamics. As $\mu$ is decreased, ion entrapment becomes sinusoidal, avoiding the sharp peaks observable in \cFIG{fig:traj}.


Using this geometry, a boundary element model was generated with Charged Particle Optics (CPO) software \cite{read:2011} to determine the electric potential on a grid of points in the region of interest around where the two traps cross. These potentials were numerically calculated for voltages applied to each trap separately (while the other trap was grounded). The output was a grid of electric fields for each trap, which could be added together after being scaled by the applied voltages and $f(t)$ to calculate the field dynamics before, during, and after the transition. To calculate ion trajectories accurately, Catmull-Rom interpolation \cite{interpol} was used to interpolate between mesh values.

For the flight simulation the state of the ion is taken to be
$
(c, \mathbf{p}, \mathbf{v})
$,
where $c, \mathbf{p}$, and $\mathbf{v}$ are the charge, position, and velocity, respectively. The ordinary differential equation governing the ion dynamics is
$$
\partial_t (c, \mathbf{p}, \mathbf{v}) = \Big(0, \mathbf{v},  \sum_{r\in rect}\nabla \Phi(\mathbf{p,t})\Big)\,,
$$
where $\sum_{r\in rect} \nabla \Phi(\mathbf{p,t})$ is the sum of the electric fields over all electrodes given predetermined voltages. 
The RK4 method was used to simulate the motion of the ion, due to its accuracy and efficiency at calculating trajectories 
for potentials defined by low-order polynomials.  



\subsection{Dynamic junction stability}\label{ssdyjust}

While the first test of junction stability using the numerical method outlined above verified ion stability at fixed voltage levels, a second  test confirmed the stability of the complete transfer protocol. The trap from \cFIG{fig:singleTrap} served as a basis for one of the layers, duplicated to produce the full local configuration, as shown within the global configuration of \cFIG{fig:dualTrap}. Initially, a spread of stable junction parameter configurations was considered, using the results from \cFIG{fig:complexMap}. 

Stable solutions were verified for $f(t) = 1/2$, demonstrating a static solution with the ion suspended between the traps for a paused junction operation. Then, transfers were validated using 
a function $f(t)$ which has three parts consisting of storing the ion with the bottom trap, linearly transferring to the upper trap, and finally storing the ion with the upper trap, all scaled to cover an arbitrary number of RF cycles. Sample paths are displayed in \cFIG{fig:traj} in conjunction with their analytic stability triplets. \commentB{The qualitative behavior of these paths is discussed in Subsection~\ref{pic-expl}}. Due to deviations in the trapping potential far from the origin, trap stability was weakened for large values of $\alpha$, to the point that stability is lost for $\alpha > 0.3$. However, as noted at the end of Sec.~\ref{sec:analyticTreatment}, large values of $\alpha$ are rarely relevant for practical ion junction operation. Additionally, \cFIG{fig:traj} illustrates differences in axial potential strength and practical configurations for ion junction tuning, as further discussed in the caption. 

Instantaneous transfers are theoretically possible, but will be practically limited by filters on the RF voltages and the need to minimize motional excitation.
In numerical tests, control electrodes successfully contained the ion, although in some transfers, the average $v_y$ was increased. This corresponds to the attenuation of the stability cross-section of the junction transfer observed as $\alpha$ increases in \cFIG{fig:complexMap}, predicting that the ion would be lost in the $y$-axis for sufficiently large $\alpha$.

\commentB{No tuning was necessary to configure the traps, as all tested trap frequencies naturally occurred in the stable region for the two-trap protocol. This is expected to hold true for common ion trap configurations. }

\subsection{\commentB{Dynamic ion path visualization}}\label{pic-expl}

\begin{figure*}[hbtp]
\includegraphics[width=\textwidth]{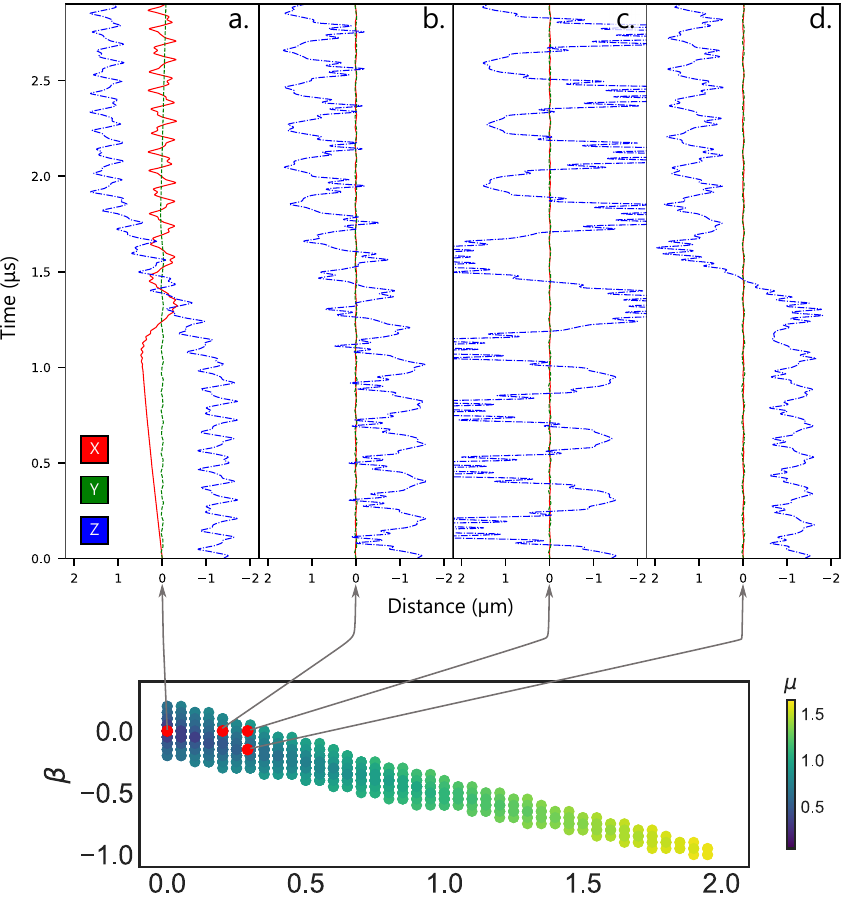}
\caption{\label{fig:traj}An illustration of stable ion flights for the two-layer junction trap with $\mu=0.75$ and varied $(\alpha, \beta) = (0,0), (0.2,0), (0.29,0), (0.29,-0.15)$ for subfigures \textbf{A--D}, respectively. The total transition time was $T=2.9\,\mu\mathrm{s}$, and $f(t)$ was $0$ until $T/3$, increased linearly to $1$ at $2T/3$, and then held there until $T$. For each trajectory, an ion at the RF null was initialized with the exaggerated initial velocity of $(5 ,5, 5)\,\mathrm{m/s}$. Red (solid), green (dashed), and blue (dash and dotted) lines correspond to $x$, $y$, and $z$ respectively.  Trajectory \textbf{A} was not predicted to be stable. The instability is confirmed by close examination of the behavior along the $y$-axis: After junction operation, the ion has a small velocity in the positive $y$-direction due to a lack of a containing $\alpha$-field. In \textbf{B--D} with $\alpha > 0$, the ion is completely stable. However, in the increase of $\alpha$ from \textbf{B} to \textbf{C}, the erosion of the $z$-axis confinement due to the $\alpha$-field generating a repulsive potential becomes increasingly evident, as the ion center is  shifted off the RF null by $\alpha$-field asymmetries. In \textbf{D}, restrengthening the $z$-axis field by making $\beta$ negative illustrates one approach to ensuring ion stability, the other being to work with small $\mu$ and $\alpha$. Values of $\alpha$ larger than $0.3$ failed to confine the ion, due to ion momentum during junction operation which was not included in our time-independent analysis.}
\end{figure*}

\commentB{To gain a holistic, qualitative portrait of the ion trajectories associated with \cFIG{fig:complexMap}, four individual ion trajectories are displayed in \cFIG{fig:traj}. The parameter $\mu$ was taken to be $0.75$, and the pairs $(\alpha, \beta)$ were chosen as $(0,0), (0.2,0), (0.29,0)$, and $(0.29,-0.15)$ for subfigures \textbf{A--D}, respectively. The total transition time was $T=2.9\,\mu\mathrm{s}$. The function $f(t)$ was $0$ until $T/3$, increased linearly to $1$ at $2T/3$, and then held at $1$ until $T$. For each trajectory, an ion at the RF null was initialized with the exaggerated initial velocity of $(5 ,5, 5)\,\mathrm{m/s}$. Red (solid), green (dashed), and blue (dash and dotted) lines correspond to $x$, $y$, and $z$ respectively. Note that as the trap junction is centered at $1.5\,\mu\mathrm{s}$, for the rotoreflected traps, the $x$- and $y$-axes are effectively switched before and after the $1.5\,\mu\mathrm{s}$ mark, and the $z$ trapping potential is inverted.}

\commentB{The first trajectory, visualized in \textbf{A}, is an example of the ion failing to remain within the confines of the trap. The reason the confinement fails along this edge of the stability regime is that the longitudinal trapping potential is non-existent, enabling the ion to drift out the two opposing ends of the trap. Before $1.5\,\mu\mathrm{s}$, the lack of a trapping potential along the $x$-axis is visible as the ion drifts to the negative direction in red. After the junction operation, the axial trapping potential confines the ion along the $x$ direction. However, as a result of the relative rotation of the traps, the ion is now no longer confined along the $y$-axis. Before the junction operation, the ion experiences fine oscillations along the $y$-axis. However, after the junction operation, these oscillations are no longer confined, manifesting as a slight bias to the right direction. This bias can be observed by noting the drift of the green line as the ion starts centered with the red oscillations, but eventually reaches the peak of the red oscillations. If the diagram were doubled in duration, this drift would continue until the ion was thrown off the end of the trap.}

\commentB{The second trajectory, visualized in \textbf{B}. is an example where the ion is successfully confined. At this point in the stability diagram, the ion is confined along both the $x$- and $y$-axes, so the only oscillation is along the $z$-axis. Here the ion maintains a smooth transition between the two trap minima at $(0,0,1\mu m)$ and $(0, 0, -1\mu m)$. Stability (in terms of ion oscillations) appears to be optimized by sampling from the center of the stability diagram.}

\commentB{The third trajectory, visualized in \textbf{C}, is an example where the ion is confined in both the $x$- and $y$-axes, but is close to the edge of the stability diagram. As this corresponds to an increase in $\alpha$, and therefore an increase in the $z$ trapping potential, the non-parabolic trapping potential is more obvious, as the ion oscillates across the center of the two traps, even while being confined.}

\commentB{Finally, \textbf{D} illustrates that the performance of \textbf{b} may be recovered by increasing $\beta$ until the trap operation is again at the center of the stability diagram. However, increased oscillations result from the increase in the trapping potential, in accord with the harmonic oscillator solution.}

\commentB{Qualitatively, this suggests that the optimal trapping potentials are found as far from the edges of the stability region as possible, and that stability degenerates as the $\mu$ value is increased.}

\section{Trap configuration and architecture}\label{sec:trapConfiguration}


In order to implement this junction scheme, the RF electrode voltage must be lowered on one entire trap while it is simultaneously raised on the other trap. There are two ways to accommodate this requirement. The first is to rely on a segmentation of the RF rails, such that particular sections of the linear traps are on while other sections are off. This nontrivial hardware change would facilitate a simple qubit transport protocol that would allow arbitrary ion movement within the array, \commentB{but would require vias and screened RF leads under the top metal layer.} Segmented RF designs have been studied and attempted, but so far with limited success due to technical challenges like equalizing the phase and amplitude of distinct but neighboring switchable RF electrodes. The second option is to use linear traps where continuous RF electrodes are all on or all off, depending on the layer. This simplifies the hardware, but limits the allowed transport at a given time to a single dimension. 

\begin{figure}[h]
\includegraphics[width=0.48\textwidth]{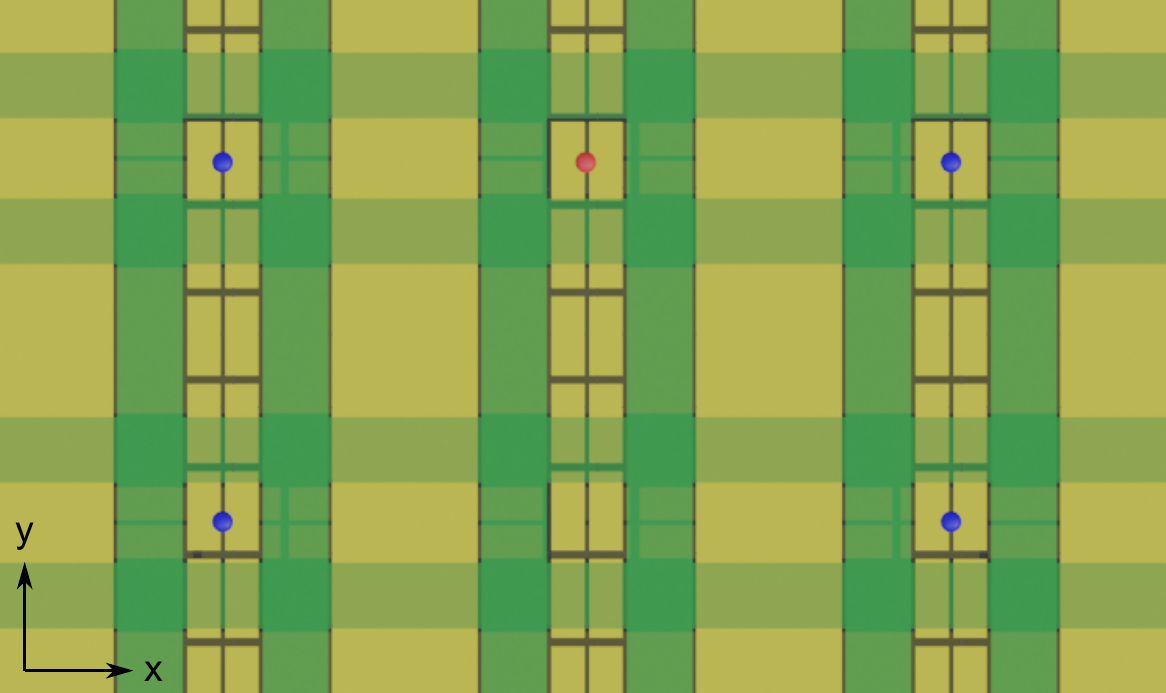}
\caption{\label{fig:globalDesign} Sample layout of four data ions (blue) and one ancilla ion (red) arranged in a surface code plaquette. In this diagram they are all trapped at the intersection of the two trapping layers, even though only one trapping layer is active at a time (except during transitions).  
}
\end{figure}

\cFIG{fig:globalDesign} presents a layout with continuous RF electrodes.  Five ions that form a plaquette in a surface code are shown; \commentB{the data ions are stored at the RF rail crossings (in $xy$) of the two traps so that they can remain laterally stationary during syndrome extraction, but move in $z$ according to which trap is enabled. The ancilla ion moves in $xy$ to interact with each neighboring data ion \cite{kang:2023}. In this scheme all ions, data and ancilla, would be confined by one layer or the other at a given time.  When making that transition they would be positioned at the RF rail crossing points and simultaneously moved in $z$.} Neighboring plaquettes with different movement patterns would require pauses in transport to accommodate times when ancilla ions are scheduled to move in orthogonal directions.
Non-nearest neighbor movement patterns are also possible with this design, and will be highly dependent on junction timing, cooling protocols, and other specifics in order to minimize latency due to transport. As an example, 
consider a particular quantum program, consisting of a compiled sequence of 
two-qubit gates entangling qubits $i$ and $j$. With a fixed maximum qubit capacity per trap, connectivity may be represented as a graph, and a simple greedy algorithm used to group qubits according to gate requirements. 


\section{Conclusion}\label{sec:conclusions}

We have introduced a novel concept for a microfabricated ion trap that supports 2D ion transport using vertically offset planar traps with RF electrodes that are perpendicularly oriented. We developed a mathematical model of our junction, and solved it analytically for bounded ion trajectories. The analytical model was augmented by a dynamic numerical simulation of successful ion transfer, 
verifying the existence of stable ion trajectories 
throughout the transfer from one trap to the other.

We highlighted two types of RF electrode architectures for synchronous ion transport to complement the unique global trap structure required by the proposed junction geometry. Segmented RF rails require modifications to the simple and continuous rail designs currently in use, but enable unrestricted ion transport protocols between traps. Continuous RF rails do not require any additional hardware design or modification, but will require coordinated transport that may lead to additional latency.

While this novel junction design solves the problem of RF lead routing for larger arrays of 2D ion traps, it introduces other challenges that have to be overcome to make it practically useful.  Some of these may be solved with additional integration, like replacing free-space delivery and collection optics with \commentB{microwave-based gates}, integrated waveguides, and detectors. A fuller analysis would also require trap-specific simulations to identify the optimal RF and control voltage protocols for transferring ions with minimal motional excitation.  The ions would not be stored at a pure RF null during the transition and therefore would experience an unavoidable amount of micromotion, so the RF drive voltage would have to be low-noise in order not cause excessive motional heating. Even with these new challenges, the analysis in this paper shows that a two-layer trap geometry based on surface traps can enable 2D qubit connectivity for trapped-ion quantum computing.\\

\section*{Data availability statement}

All data that support the findings of this study are included within the article (and any supplementary files).\\

\section*{Acknowledgements}

This research was conducted at Ames National Laboratory for the U.S. DOE with Iowa State University under Contract No. DE-AC02–07CH11358. This material is also based upon work supported by the U.S. Department of Energy, Office of Science, National Quantum Information Science Research Centers, Quantum Systems Accelerator. Sandia National Laboratories is a multimission laboratory managed and operated by National Technology \& Engineering Solutions of Sandia, LLC, a wholly owned subsidiary of Honeywell International Inc., for the U.S. Department of Energy’s National Nuclear Security Administration under contract DE-NA0003525. This paper describes objective technical results and 
analysis. Any subjective views or opinions that might be expressed in the paper do not necessarily represent the views of the U.S. Department of Energy or the United States Government.




\bibliography{citations}

\end{document}